\documentclass[10pt]{article}
\usepackage{amsmath}
\usepackage{braket}
\usepackage[OE]{express}

\begin{document}

\title{Longer distance continuous variable quantum key distribution protocol with photon subtraction at receiver}

\author{Kyongchun Lim, Changho Suh, and June-Koo Kevin Rhee\authormark{*}}

\address{School of Electrical Engineering, Korea Advanced Institute of Science and Technology (KAIST),  291 Daehak-ro,
Yuseong-gu, Daejeon 34141, Republic of Korea}

\email{\authormark{*}jk.rhee@kaist.ac.kr} 



\begin{abstract}
One of the limitation of continuous variable quantum key distribution is relatively short transmission distance of secure keys. In order to overcome the limitation, some solutions have been proposed such as reverse reconciliation, trusted noise concept, and non-Gaussian operation. In this paper, we propose a protocol using photon subtraction at receiver which utilizes synergy of the aforementioned properties. By simulations, we show performance of the proposed protocol outperforms other conventional protocols. We also find the protocol is more efficient in a practical case. Finally, we provide a guide for provisioning a system based on the protocol through an analysis for noise from a channel.
\end{abstract}


\bibliographystyle{osajnl}
\bibliography{mybib}

\section{Introduction}
Quantum key distribution (QKD) is one of realistic applications of quantum technologies. Technically, QKD provides shared secret keys between two remote parties, Alice and Bob. Because of its nature of unconditional security, QKD has attracted broad research interest since the first QKD protocol, the BB84 protocol, was  invented in 1984 \cite{bennett1984quantum}. In early days, researches about discrete variable QKD (DVQKD) had been actively conducted. After a while, continuous variable QKD (CVQKD) encoding secret key information on continuous degree of freedom in phase space of a quantum state started gaining interest triggered by the proposal by \cite{PhysRevA.61.010303}. Unlike DVQKD, CVQKD is beneficial to generate high secure key rates at short distances due to high dimensionality of continuous variable, but it could not achieve key sharing over a long distance. One of main hurdles of the limited distance is caused from low error correction efficiency at a long distance under Gaussian based CVQKD protocols \cite{PhysRevA.77.042325}. In order to overcome the hurdle, a variety of ways in terms of post-processing are proposed such as post-selection\cite{PhysRevLett.89.167901}, multi-dimensional reconciliation\cite{PhysRevA.77.042325}, and reverse reconciliation\cite{grosshans2003quantum}. Especially, the reverse reconciliation can overcome fairly the limit by a simple way changing reference of error correction, which provides importance of a receiver side in CVQKD protocols. There are also many proposals to modify a quantum channel. Some proposals utilize amplifiers including noiseless amplifier\cite{PhysRevA.86.060302, zhang2014improvement, e17074547}, while others change transmitter to use four-state or eight-state Gaussian states\cite{xuan200924, 2058-9565-2-2-024010, 0256-307X-30-1-010305}. As another way to increase distance, trusted noise concept is proposed, with which a noise added in a proper way can increase security \cite{usenko2016trusted}. For example, noise added by a receiver, Bob, can increase security in a CVQKD protocol with reverse reconciliation.

Deviated from Gaussian state protocols, non-Gaussian state protocols have been researched due to its potential for high security in terms of secure key rate and distance\cite{borelli2016quantum, PhysRevA.83.042312, PhysRevA.87.012317, PhysRevA.93.012310}. One of the representative non-Gaussian operation used in a CVQKD protocol is photon subtraction implemented with a beam splitter and a photon counter. This shows that distance can be improved with photon subtracted states.

In this paper, we propose a CVQKD protocol utilizing the aforementioned properties. Specifically, the protocol is a reverse reconciliation based CVQKD protocol adopting photon subtraction operation in a receiver side. Through numerical simulations, we show the proposed protocol outperforms a conventional CVQKD protocols. This result coincides with existed research results showing positive potentials of non-Gaussian state and trusted noise in terms of security, which convinces our approach is quite reasonable.

The paper is organized as we introduces a target system model in Section II. A secure key rate based on the model is obtained in Section III. Section IV provides numerical simulations for secure key rates under the model. We finalize in Section V with some concluding remarks.

\section{System Model}

\begin{figure}[!t]
\centering

{\includegraphics[height=3cm]{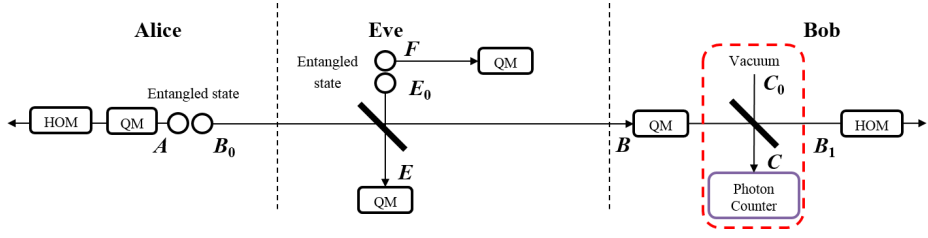}}~

\caption{A model for CVQKD with photon subtraction under a collective attack. HOM and QM stand for homodyne detection and quantum memory, respectively.}

\label{model}
\end{figure}

In this section, we introduce a system model for a CVQKD protocol with photon subtraction at a receiver, Bob, as shown in Fig. \ref{model}. Here, a transmitter, Alice prepares a two mode squeezed vacuum (TMSV) state $\ket{\psi}_{AB_0}$ which is an entangled state in a continuous variable domain.
\begin{align}
\ket{\psi}_{AB_0} = \sum_{n=0}^{\infty}{\alpha_n\ket{n,n}_{AB_0}},
\end{align}
where 
\begin{align}
\alpha_n = \sqrt{\frac{\alpha^{2n}}{\left(1+\alpha^2\right)^{n+1}}},
\end{align}
$\alpha^2$ and $\ket{n}$ represent mean photon number and number state, respectively.

In this model, we assume a collective attack where an Eve's initial state is prepared as a TMSV state $\ket{\psi}_{E_0F}$. In a similar to $\ket{\psi}_{AB_0}$, $\ket{\psi}_{E_0F}$ can be expressed with a mean photon number of noise from a channel, $\beta^2$.
\begin{align}
\ket{\psi}_{E_0F} = \sum_{m=0}^{\infty}{\beta_m\ket{m,m}_{E_0F}},
\end{align}
where 
\begin{align}
\beta_m = \sqrt{\frac{\beta^{2m}}{\left(1+\beta^2\right)^{m+1}}}
\end{align}
After preparing $\ket{\psi}_{AB_0}$, she transmits a quantum state $\rho_{B_0} = \mathrm{Tr}_A \left(\ket{\psi}\bra{\psi}_{AB_0} \right)$ of $\ket{\psi}_{AB_0}$ to Bob through a quantum channel, while the other quantum state $\rho_A = \mathrm{Tr}_{B_0} \left(\ket{\psi}\bra{\psi}_{AB_0} \right)$ is kept a quantum memory. The transmitted $\rho_{B_0}$ is mixed with an Eve's state $\rho_{E_0} = \mathrm{Tr}_F \left(\ket{\psi}\bra{\psi}_{E_0F} \right)$ in the channel which can be modeled as a beam splitter with transmittance $T$.

Before finding a quantum state after a channel, we first look into a principle of a beam splitter. Assume there is a beam splitter characterized transmittance $T$, where there are two input creation operators $\hat{i}^\dag_{1}$, $\hat{i}^\dag_{2}$ and two output creation operators $\hat{o}^\dag_{1}$, $\hat{o}^\dag_{2}$. In this setup, output operators can be expressed based on the input operators as follows:
\begin{align}
\hat{o}^\dag_{1} = \sqrt{T}\hat{i}^\dag_{1} + \sqrt{1-T}\hat{i}^\dag_{2}, \\
\hat{o}^\dag_{2} = \sqrt{T}\hat{i}^\dag_{2} - \sqrt{1-T}\hat{i}^\dag_{1}.
\end{align}
By rearranging the above equations,
\begin{align}
\hat{i}^\dag_{1} = \sqrt{T}\hat{o}^\dag_{1} - \sqrt{1-T}\hat{o}^\dag_{2}, \\
\hat{i}^\dag_{2} = \sqrt{T}\hat{o}^\dag_{2} + \sqrt{1-T}\hat{o}^\dag_{1}.
\end{align}

\noindent Now, we can find a quantum state after a channel $\ket{\psi}_{AB_1EF}$ based on the initial states of Alice and Eve by applying a beam splitter operation on them.
\begin{align}
\ket{\psi}_{AB_0E_0FB_1E} &= \sum^{\infty}_{n=0}{\alpha_n\ket{n,n}_{AB_0}} \sum^{\infty}_{m=0}{\beta_m\ket{m,m}_{E_0F}} \ket{0,0}_{B_1E}, \\
\label{creation_operator}
&= \sum^{\infty}_{n=0}{\alpha_n\frac{\left(\hat{b}^\dag_0\right)^n}{\sqrt{n!}}\ket{n,0}_{AB_0}} \sum^{\infty}_{m=0}{\beta_m\frac{\left(\hat{e}^\dag_0\right)^n}{\sqrt{m!}}\ket{0,m}_{E_0F}} \ket{0,0}_{B_1E}, \\
\label{BS_operation}
&= \sum^{\infty}_{n=0}{\alpha_n\frac{\left(\sqrt{T}\hat{b}^\dag_1 - \sqrt{1-T}\hat{e}^\dag\right)^n}{\sqrt{n!}}\ket{n,0}_{AB_0}} \\ 
&= \sum^{\infty}_{n=0}{\alpha_n\sum^{n}_{k=0}{(-1)^k\sqrt{
n \choose k
}\left(\sqrt{T} \right)^{n-k} \left(\sqrt{1-T} \right)^{k}}\ket{n,0}_{AB_0}} \\
&\hspace{2mm}\otimes \sum^{\infty}_{m=0}{\beta_m\sum^{m}_{l=0}{\sqrt{
m \choose l
}\left(\sqrt{T} \right)^{m-l} \left(\sqrt{1-T} \right)^{l}}\sqrt{
n-k+l \choose l
}\sqrt{
k+m-l \choose k
}}\nonumber \\
&\hspace{2mm} \ket{0,m}_{E_0F}\ket{n-k+l,k+m-l}_{B_1E}, \nonumber
\end{align}
where $\hat{a}^\dag$ indicates a creation operator of a quantum state $\rho_A$. Eq. (\ref{creation_operator}) comes from a relation between a number state and a creation operator such that $\hat{a}^\dag \ket{n} = \sqrt{n+1}\ket{n+1}$. A beam splitter operation results in Eq. (\ref{BS_operation}). We can easily obtain $\ket{\psi}_{AB_1EF}$ by tracing out and rearranging $\ket{\psi}_{AB_0E_0FB_1E}$.
\begin{align}
\ket{\psi}_{AB_1EF} = \sum^{\infty}_{n=0}{\alpha_n\sum^{n}_{k=0}{(-1)^k \gamma^T_{n,k}}}  \sum^{\infty}_{m=0}{\beta_m\sum^{m}_{l=0}{\gamma^T_{m,l}}} \zeta_{n,k,m,l} \ket{n,n-k+l,k+m-l,m}_{AB_1EF},
\end{align}
where
\begin{align}
\gamma^{T}_{n,k} &= \sqrt{n \choose k}\left(\sqrt{T} \right)^{n-k}\left(\sqrt{1-T} \right)^{k}, \\
\zeta_{n,k,m,l} &= \sqrt{n-k+l \choose l}\sqrt{k+m-l \choose k}.
\end{align}

In a similar way, a quantum state after photon subtraction represented as a beam splitter with transmittance $T_1$ $\ket{\psi}_{AB_2EFC}$ can be obtained as follows:
\begin{align}
\label{PS_state}
\ket{\psi}_{AB_2EFC} =&
\sum^{\infty}_{n=0}{\alpha_n\sum^{n}_{k=0}{(-1)^k \gamma^T_{n,k}}}  \sum^{\infty}_{m=0}{\beta_m\sum^{m}_{l=0}{\gamma^T_{m,l}}} \zeta_{n,k,m,l} \sum^{n-k+l}_{s=0} (-1)^s \gamma^{T_1}_{n-k+l,s} \\ &\ket{n,n-k+l-s,k+m-l,m,s}_{AB_2EFC}. \nonumber
\end{align}
For easy understanding, we first analyze single photon subtracted case corresponding to $C$ is a single photon state. Then, the photon subtracted state $\ket{\psi}\bra{\psi}^{\textrm{PS}}_{AB_2EF} = \textrm{Tr}_C\left( \ket{\psi}\bra{\psi}^{\textrm{PS}}_{AB_2EFC}|_{s=1}\right)$ has the following expression.
\begin{align}
\ket{\psi}^{\textrm{PS}}_{AB_2EF} =&
-\frac{1}{\sqrt{P_1}}\sum^{\infty}_{n=0}{\alpha_n\sum^{n}_{k=0}{(-1)^k \gamma^T_{n,k}}}  \sum^{\infty}_{m=0}{\beta_m\sum^{m}_{l=0}{\gamma^T_{m,l}}} \zeta_{n,k,m,l} \gamma^{T_1}_{n-k+l,1} \\ &\ket{n,n-k+l-1,k+m-l,m}_{AB_2EF}, \nonumber
\end{align}
where $P_1$ refers a normalization parameter defined as a probability that $C$ is in a single photon state.
\begin{align}
P_1 &= \braket{\psi|\psi}^{\textrm{PS}}_{AB_2EF} \\
&= \sum^{\infty}_{n=0} \alpha^2_n \sum^{n}_{k=0} \sum^{\infty}_{m=0} \beta^2_m \sum^m_{l=0} \left(J^{+}_{(n,k,m,l),(n,k,m,l)} + J^{-}_{(n,k,m,l)(n,k,m,l)} \right),
\end{align}
where
\begin{align}
&J^{+}_{(n_1,k_1,m_1,l_1)(n_2,k_2,m_2,l_2)} =\\
&\sum^{\min\{n_2-k_2,m_2-l_2\}}_{j=0} (-1)^j \gamma^T_{n_1,k_1} \gamma^T_{n_2,k_2+j} \gamma^T_{m_1,l_1} \gamma^T_{m_2,l_2+j} \zeta_{n_1,k_1,m_1,l_1} \zeta_{n_2,k_2+j,m_2,l_2+j} \gamma^{T_1}_{n_1-k_1+l_1,1} \gamma^{T_1}_{n_2-k_2+l_2,1}, \nonumber \\
&J^{-}_{(n_1,k_1,m_1,l_1)(n_2,k_2,m_2,l_2)} =\\
&\sum^{\min\{k_2,l_2\}}_{j=1} (-1)^j \gamma^T_{n_1,k_1} \gamma^T_{n_2,k_2-j} \gamma^T_{m_1,l_1} \gamma^T_{m_2,l_2-j} \zeta_{n_1,k_1,m_1,l_1} \zeta_{n_2,k_2-j,m_2,l_2-j} \gamma^{T_1}_{n_1-k_1+l_1,1} \gamma^{T_1}_{n_2-k_2+l_2,1}. \nonumber
\end{align}
Finally, we obtain a whole quantum state after photon subtraction $\ket{\psi}^{\textrm{PS}}_{AB_2EF}$ which is not a Gaussian state anymore. Then, $\ket{\psi}^{\textrm{PS}}_{AB_2EF}$ is measured by a homodyne detector with respect to $q$ or $p$ quadrature. After a choice of the quadrature is publicly announced to Alice, the kept $\rho_A$ in a quantum memory is measured by homodyne detector with respect to the quadrature. In succession, Alice and Bob perform reverse reconciliation and privacy amplification to generate final secure keys.


\section{Secure key rate}
\begin{figure}[!t]
\centering

{\includegraphics[height=3cm]{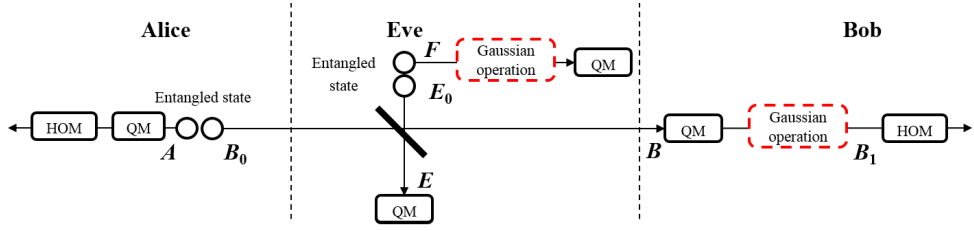}}~

\caption{A model for CVQKD substituting photon subtraction to a Gaussian unitary operation resulting in the same covariance under a collective attack. }

\label{model_eq}
\end{figure}

In this section, we calculate secure key rate under our system model. Under a collective attack, secure key rate $K$ can be calculated as follows:
\begin{align}
K = P_1 \left\{ \beta I(A:B_2) - \chi(B_2:EF) \right\},
\end{align}
where $\beta$ is reconciliation efficiency. Here, $I(A:B_2)$ represents mutual information between Alice and Bob, while $\chi(B_2:EF)$ refers the Holevo information defined as the maximum information extracted by Eve from Bob's data. Calculating $I(A:B_2)$ and $\chi(B_2:EF)$ starts from a photon subtracted state with a density matrix $\rho^{\textrm{PS}}_{AB_2EF} = \ket{\psi} \bra{\psi}^{\textrm{PS}}_{AB_2EF}$. In case of $\chi(B_2:EF)$, it can be calculated by
\begin{align}
\label{Holevo}
\chi(B_2:EF) &= S(\rho^{\textrm{PS}}_{EF}) - S(\rho^{\textrm{PS}}_{EF|B_2}), \\
&= -\sum_{i} \lambda^{EF}_i \log_2{\lambda^{EF}_i} +\sum_{i} \lambda^{EF|B_2}_i \log_2{\lambda^{EF|B_2}_i},
\end{align}
where $S(\cdot)$ represents von Neumann entropy. Here, a density matrix for an Eve's state is denoted as $\rho^{\textrm{PS}}_{EF}$ having eigenvalues $\{\lambda^{EF}_i\}$, while a density matrix for an Eve's state given Bob's measurement on $B_2$ is represented as $\rho^{\textrm{PS}}_{EF|B_2}$ with eigenvalues $\{\lambda^{EF|B_2}_i\}$. Note that there are infinite number of $\{\lambda^{EF}_i\}$ and $\{\lambda^{EF|B_2}_i\}$ due to infinite dimension of $\rho^{\textrm{PS}}_{EF}$ and $\rho^{\textrm{PS}}_{EF|B_2}$. Therefore, infinite eigenvalues and their infinite sum make calculation of $\chi(B_2:EF)$ intractable. 

If we remove one of the two calculations, the intractability of $\chi(B_2:EF)$ can be improved. Removing calculation of infinite eigenvalues can be done if all states are Gaussian states because $S(\cdot)$ of a Gaussian state can be calculated based on its covariance matrix having finite dimension. In order to do that, we substitute photon subtraction operation to a Gaussian unitary operation making the same covariance matrix with ${\Gamma}^{\textrm{PS}}_{AB_2EF}$ of $\rho^{\textrm{PS}}_{AB_2EF}$ as in Fig. \ref{model_eq}. This provides a lower bound of the original protocol in terms of performance by the theorem of Gaussian optimality \cite{garcia2006unconditional}. Define a state made by the Gaussian unitary operation as $\rho^{\textrm{G}}_{AB_2EF}$ with ${\Gamma}^{\textrm{G}}_{AB_2EF} = {\Gamma}^{\textrm{PS}}_{AB_2EF}$. Then, by the theorem,
\begin{align}
K \geq P_1 \left\{ \beta I_G(A:B_2) - \chi_G(B_2:EF) \right\},
\end{align}
where $I_G(A:B_2)$ and $\chi_G(B_2:EF)$ represent mutual information between Alice and Bob and the Holevo information obtained from $\rho^{\textrm{G}}_{AB_2EF}$, respectively. Since $\rho^{\textrm{G}}_{AB_2EF}$ is a Gaussian state, $I_G(A:B_2)$ and $\chi_G(B_2:EF)$ can be easily obtained from a covariance matrix of ${\Gamma}^{\textrm{G}}_{AB_2EF}$.

In order to calculate ${\Gamma}^{\textrm{PS}}_{AB_2EF}$, we first look into a structure of a covariance matrix. Define a operator vector $\hat{x}$ as follows:
\begin{align}
\hat{x} = \left( \hat{q}_1, \hat{p}_1,..., \hat{q}_N, \hat{p}_N \right)^{\top},
\end{align}
where $\hat{q}_i$ and $\hat{p}_i$ are quadrature operators of the $i$-th mode out of $N$ modes. Then, an element of a covariance matrix ${\Gamma}$ is defined as
\begin{align}
\label{covariance}
\Gamma_{ij} = \frac{1}{2}\braket{\Delta\hat{x}_i\Delta\hat{x}_j + \Delta\hat{x}_j\Delta\hat{x}_i},
\end{align}
where $\Delta\hat{x}_i := \hat{x}_i - \braket{\hat{x}_i}$. In general, $\Gamma_{ij}$ indicates a correlation between modes $i$ and $j$, while a diagonal element $\Gamma_{ii}$ indicates a variance of a mode $i$. From now on, for easy understanding, we substitute $\Gamma_{ii}$ and $\Gamma_{ij}$ to $V_{i}$ and $C_{ij}$, respectively.

Since $\rho^{\textrm{G}}_{AB_2EF}$ is a Gaussian state, $I_G(A:B_2)$ is calculated as follows:
\begin{align}
I_G(A:B_2) = \frac{1}{2} \log_2{\frac{V_{B_2}}{V_{B_2|A}}},
\end{align}
where $V_{B_2|A}$ is a conditional variance of Bob's data given Alice's data. Since $q$ and $p$ quadrature operators are independent and symmetric in terms of probability distribution, $V_{B_2}$ is invariant with respect to the quadratures.
\begin{align}
V_{B_2} &= \frac{1}{2}\braket{\Delta\hat{q}_{B_2}\Delta\hat{q}_{B_2} + \Delta\hat{q}_{B_2}\Delta\hat{q}_{B_2}}, \\
&= \braket{\psi|1+2\hat{b}^\dag_2\hat{b}_2|\psi}^{\textrm{PS}}_{AB_2EF}, \\
\label{V_B2}
&= 1 + \frac{2}{P_1}\sum^{\infty}_{n=0} \alpha^2_n \sum^{n}_{k=0} \sum^{\infty}_{m=0} \beta^2_m \sum^m_{l=0} (n-k+l-1) \left( J^{+}_{(n,k,m,l),(n,k,m,l)} + J^{-}_{(n,k,m,l)(n,k,m,l)} \right)   \\ &\hspace{2mm} \times u(n-k+l-1),\nonumber
\end{align}
where $u(\cdot)$ is a step function defined as
\begin{align}
u(x) = 
\begin{cases}
1,& \textrm{if $x \geq 1$} \\
0,& \textrm{otherwise}.
\end{cases}
\end{align}
The second equality holds since $\braket{\hat{q}_{B_2}}=0$ and the relations that $\hat{q} = \hat{b}^\dag_2 + \hat{b}_2$ and $\hat{b}_2\hat{b}^\dag_2 - \hat{b}^\dag_2\hat{b}_2 = 1$.

In case of $V_{B_2|A}$ where $B_2$ given $A$ follows a Gaussian distribution, this can be obtained as follows:
\begin{align}
V_{B_2|A} = V_{B_2} - \frac{C_{AB_2}}{V_A},
\end{align}
where
\begin{align}
V_A = 1 + \frac{2}{P_1}\sum^{\infty}_{n=0} \alpha^2_n \sum^{n}_{k=0} \sum^{\infty}_{m=0} \beta^2_m \sum^m_{l=0} n \left( J^{+}_{(n,k,m,l),(n,k,m,l)} + J^{-}_{(n,k,m,l)(n,k,m,l)} \right),
\end{align}
and
\begin{align}
C_{AB_2} &= \frac{1}{2}\braket{\Delta\hat{q}_{A}\Delta\hat{q}_{B_2} + \Delta\hat{q}_{B_2}\Delta\hat{q}_{A}}, \\
&= \frac{2}{P_1}\sum^{\infty}_{n=0} \alpha_n \alpha_{n+1} \sum^{n}_{k=0} \sum^{\infty}_{m=0} \beta^2_m \sum^m_{l=0} \sqrt{n+1} \sqrt{n-k+l} \\
&\hspace{2mm}\times \left( J^{+}_{(n,k,m,l),(n+1,k,m,l)} + J^{-}_{(n,k,m,l)(n+1,k,m,l)} \right). \nonumber
\end{align}

In the case of Eve's information, for a Gaussian state, the Holevo information becomes 
\begin{align}
\label{holevo}
\chi_G(B_2:EF) = \sum_i g(v^{EF}_i) - \sum_j g(v^{EF|B_2}_j),
\end{align}
where 
\begin{align}
g(x) = \left( \frac{x+1}{2} \right) \log_2{\left( \frac{x+1}{2} \right)} - \left( \frac{x-1}{2} \right) \log_2{\left( \frac{x-1}{2} \right)}
\end{align}
Here, $v^{EF}_i$ and $v^{EF|B_2}_j$ indicate an $i$-th symplectic eigenvalue of a covariance matrix for Eve's state ${\Gamma}^{G}_{EF}$ and a $j$-th symplectic eigenvalue of a covariance matrix for Eve's state given Bob's measurement ${\Gamma}^{G}_{EF|B_2}$, respectively. Specifically, $v^{EF}_i$ and $v^{EF|B_2}_i$ are calculated as absolute eigenvalues of $\sqrt{-1}\Omega {\Gamma}^G_{EF}$ and $\sqrt{-1}\Omega {\Gamma}^G_{EF|B_2}$, where $\Omega = \bigoplus_{i=1}^N \begin{bmatrix}
0 & 1 \\ -1 & 0
\end{bmatrix}$. For the symplectic eigenvalues, it is required to find structures of ${\Gamma}^G_{EF}$ and ${\Gamma}^G_{EF|B_2}$, which are obtained from ${\Gamma}^G_{EFB_2}$. Elements of ${\Gamma}^G_{EFB_2}$ can be directly calculated by using Eq. (\ref{covariance}), which provides the following form.
\begin{align}
\label{gamma_EFB2}
{\Gamma}^G_{EFB_2} = 
\begin{bmatrix}
{\Gamma}^G_{EF} & {\bf L}_{EFB_2} \\
{\bf L}_{EFB_2} & {\Gamma}^G_{B_2}
\end{bmatrix},
\end{align}
where
\begin{align}
\label{gamma__G_EF}
{\Gamma}^G_{EF} &= 
\begin{bmatrix}
V_E {\bf I}_2 & C_{EF} \sigma_z \\
C_{EF} \sigma_z & V_F {\bf I}_2
\end{bmatrix}, \\
{\bf L}_{EFB_2} &= 
\begin{bmatrix}
C_{EB_2} {\bf I}_2 \\
C_{FB_2} \sigma_z
\end{bmatrix},
\end{align}
where ${\bf I}_2$ and $\sigma_z$ represent two dimensional identity matrix and pauli $z$ operator, respectively. ${\Gamma}^G_{EF}$ is directly obtained from ${\Gamma}^G_{EFB_2}$, while ${\Gamma}^G_{EF|B_2}$ is obtained from it by using the following relation\cite{weedbrook2012gaussian}. 
\begin{align}
{\Gamma}^G_{EF|B_2} = {\Gamma}^G_{EF} - {\bf L}_{EFB_2} \left( \Pi {\Gamma}^G_{B_2} \Pi \right)^{\textrm{MP}} {\bf L}^{\top}_{EFB_2},
\end{align}
where MP refers Moore-Penrose pseudoinverse.

Note that Eve can also perform a Gaussian operation on her states as shown in Fig. \ref{model_eq} to obtain a Gaussian state with $\Gamma^{\textrm{PS}}_{EFB_2}$ because Bob performs a Gaussian operation for substituting photon subtraction. Based on this, elements of the covariance matrix in Eq. (\ref{gamma__G_EF}) are as follows:
\begin{align}
\label{Eve_strategy_w_operation_start}
V_E &= 1 + \frac{2}{P_1}\sum^{\infty}_{n=0} \alpha^2_n \sum^{n}_{k=0} \sum^{\infty}_{m=0} \beta^2_m \sum^m_{l=0} (k+m-l) \left( J^{+}_{(n,k,m,l),(n,k,m,l)} + J^{-}_{(n,k,m,l)(n,k,m,l)} \right),\\
V_F &= 1 + \frac{2}{P_1}\sum^{\infty}_{n=0} \alpha^2_n \sum^{n}_{k=0} \sum^{\infty}_{m=0} \beta^2_m \sum^m_{l=0} m \left( J^{+}_{(n,k,m,l),(n,k,m,l)} + J^{-}_{(n,k,m,l)(n,k,m,l)} \right),\\
C_{EF} &= \frac{2}{P_1}\sum^{\infty}_{n=0} \alpha^2_n \sum^{n}_{k=0} \sum^{\infty}_{m=0} \beta_m \beta_{m+1} \sum^m_{l=0} \sqrt{m+1} \sqrt{k+m-l+1} \\
&\hspace{2mm}\times \left( J^{+}_{(n,k,m,l),(n,k,m+1,l)} + J^{-}_{(n,k,m,l)(n,k,m+1,l)} \right), \nonumber \\
C_{EB_2} &= \frac{2}{P_1}\sum^{\infty}_{n=0} \alpha^2_n \sum^{n}_{k=0} \sum^{\infty}_{m=0} \beta^2_m \sum^m_{l=0} \sqrt{n-k+l} \sqrt{k+m-l} \\
&\hspace{2mm}\times \left( J^{+}_{(n,k,m,l),(n,k,m,l+1)} + J^{-}_{(n,k,m,l)(n,k,m,l+1)} \right),\nonumber\\
C_{FB_2} &= \frac{2}{P_1}\sum^{\infty}_{n=0} \alpha^2_n \sum^{n}_{k=0} \sum^{\infty}_{m=0} \beta_m \beta_{m+1} \sum^m_{l=0} \sqrt{m+1} \sqrt{n-k+l} \label{Eve_strategy_w_operation_end} \\
&\hspace{2mm}\times \left( J^{+}_{(n,k,m,l),(n,k,m+1,l+1)} + J^{-}_{(n,k,m,l)(n,k,m+1,l+1)} \right).\nonumber
\end{align}
Eqs. (\ref{holevo}) and (\ref{gamma_EFB2}) based on the above equations provide the corresponding Holevo information.

The aforementioned analysis deals with one photon subtraction case. In a similar way, general photon subtraction cases are also analyzed by adjusting the summation of $s$ in Eq. (\ref{PS_state}). Furthermore, instead of a photon counter for photon subtraction, a photon detector case can also be analyzed. This case is analyzed by summing $s$ from 1 to $\infty$ in Eq. (\ref{PS_state}).
\begin{align}
\label{PS_state_PD}
\ket{\psi}_{AB_2EFC} =&
\sum^{\infty}_{n=0}{\alpha_n\sum^{n}_{k=0}{(-1)^k \gamma^T_{n,k}}}  \sum^{\infty}_{m=0}{\beta_m\sum^{m}_{l=0}{\gamma^T_{m,l}}} \zeta_{n,k,m,l} \sum^{n-k+l}_{s=1} (-1)^s \gamma^{T_1}_{n-k+l,s} \\ &\ket{n,n-k+l-s,k+m-l,m,s}_{AB_2EFC}. \nonumber
\end{align}

\begin{figure}[t]
\centering
{
\includegraphics[height=8cm]{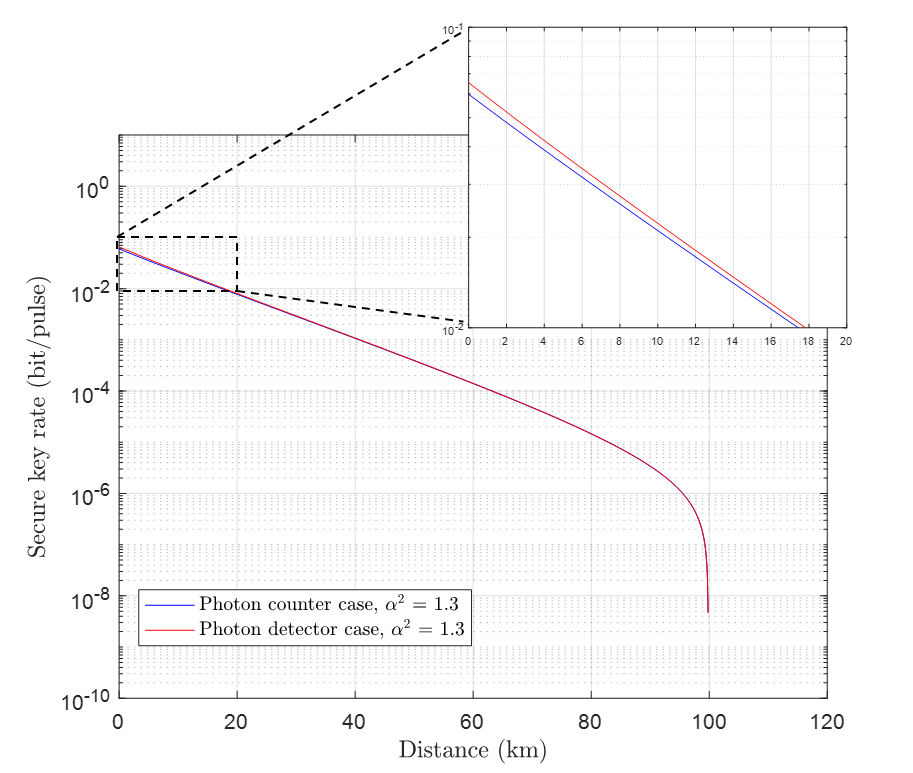}
\caption{Comparisons for secure key rate depending on devices used for photon subtraction.}
\label{result1}
}
\end{figure}

\begin{figure}[!t]
\centering
{
\includegraphics[height=8cm]{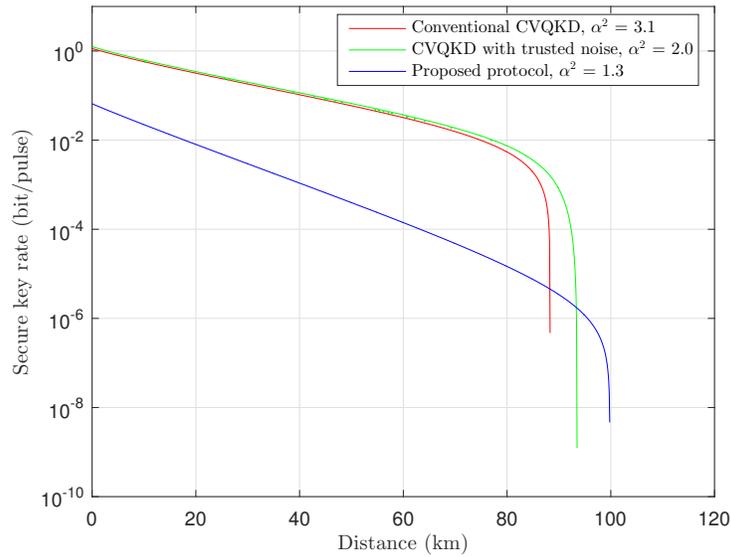}
\caption{Comparisons for secure key rate of CVQKD protocols. The mean photon number $\alpha^2$ is optimized for each case to have the maximum secure key rate.}
\label{result2}
}
\end{figure}

\section{Simulation results}

In this section, we perform three simulations. In order to check feasibility of the proposed protocol, we compare the performances of the cases for photon detector and photon counter. Based on the results of the first simulation, we compare the performance of the proposed protocol with other conventional protocols. Finally, we simulate the maximum transmission distances for protocols with respect to mean photon number of noise from a channel to analyze an effective region of the proposed protocol.

 The proposed protocol is initially based on a photon counter for photon subtraction. However, a sophisticated photon counter is hard to implement and expensive. On the other hand, a photon detector with no photon number resolution is relatively easy to implement and costs lower. For a more practical analysis, we compare two cases for a photon counter and a photon detector used in photon subtraction. Here, a photon counter case corresponds to a case for one photon subtraction case. We set error correction efficiency $f$, detector efficiency, mean photon number of noise from a channel $\beta^2$, and transmittance $T_1$ of the beam splitter of Bob as 0.95, 0.68, 0.001, and 0.9, respectively. Here, Alice's modulation variance of each case is optimized for the distance. Note that the expressions for results require infinite sums. For simulation, we truncate marginal portion of the summations by setting infinity to 30 in the summations. The corresponding results are shown in Fig. \ref{result1}. The blue and red lines indicate cases for photon counter and photon detector, respectively. From the results, we can identify a photon detector case is slightly more secure than that of a photon counter. This provides the proposed protocol is rather more secure even for a practical case. Furthermore, similar trend in the results means one photon subtraction occupies a dominant part of the performance in a photon detector case.

Next, we conduct performance comparisons between the proposed protocol with a photon detector and other conventional protocols. Here, we set simulation parameters as the same as the first simulation. For performance comparison, we simulate secure key rates of a conventional CVQKD corresponding a CVQKD without photon subtraction and a CVQKD proposed in \cite{ottaviani2016secret} where the trusted noise concept is used for better security. The corresponding results are plotted in Fig. \ref{result2} where mean photon number of Alice for each protocol is optimized. Red, green, and blue lines indicate a conventional CVQKD, a CVQKD with trusted noise, and the proposed protocol, respectively. From the results, we find that our proposed protocol can transmit secure keys at a longer distance. This yields that a combination of reverse reconciliation, the trusted noise concept, and non-Gaussian operation shows positive synergy in terms of security than that using partial properties.

\begin{figure}[!t]
\centering
{
\includegraphics[height=8cm]{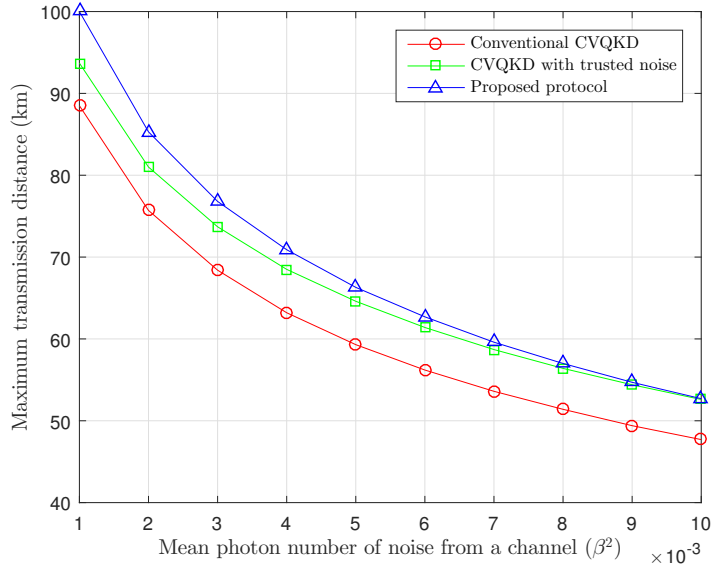}
\caption{Comparisons for the maximum transmission distance with respect to mean photon number of noise from a channel depending on CVQKD protocols.}
\label{result3}
}
\end{figure}

Finally, we conduct simulations for the maximum transmission distances for the protocols with respect to mean photon number of noise from a channel $\beta^2$. Here, the maximum distance is defined as the maximum distance with a positive secure key rate. Simulation parameters are the same as in the aforementioned simulations. As shown results in Fig. \ref{result3}, performance improvement of the proposed protocol decreases as $\beta^2$ increases, which means performance improvement of the proposed protocol is bounded in a certain channel noise region. Based on this as a guide for system provisioning, we utilize the proposed protocol depending on conditions of a system.

\section{Conclusion}
In this paper, we investigate how to improve secure distance of CVQKD. We utilize three properties of non-Gaussian state in terms of security, reverse reconciliation, and trusted noise concept to come up with a reverse reconciliation based CVQKD protocol with a photon subtracted states at a receiver. In order to overcome intractability of a calculation for secure key rate, we utilize the Gaussian optimality theorem, which yields a lower bound of secure key rate under the proposed protocol. Through simulation results, we find that the proposed protocol can improve secure distance. The proposed protocol also outperforms even for a practical case where a photon detector instead of a photon counter is used. The results convince that a combination of the properties shows positive synergy in terms of security. Furthermore, we find an effective region where the proposed protocol outperforms so that this provides a guide for provisioning of a CVQKD system based on the proposed protocol.

\section*{Funding}
ICT R\&D program of MSIP/IITP (1711057505, Reliable crypto-system standards and core technology development for secure quantum key distribution network)

\section*{Acknowledgement}
The authors would like to thank Korea Advanced Institute of Science and Technology (KAIST), Republic of Korea for supporting this work.
\end{document}